\documentclass[sigconf]{acmart}

\usepackage{enumitem}
\usepackage[most]{tcolorbox}
\usepackage{makecell}
\usepackage{booktabs}
\usepackage{multirow}
\usepackage[normalem]{ulem}
\usepackage{color}

\usepackage[ruled,vlined,linesnumbered]{algorithm2e}
\DontPrintSemicolon
\SetKwInOut{KwIn}{Input}      
\SetKwInOut{KwOut}{Output}    
\usepackage{array}
\usepackage[dvipsnames]{xcolor}
\useunder{\uline}{\ul}{}
\AtBeginDocument{%
  }
\usepackage[caption=false,font=footnotesize]{subfig}
\setcopyright{acmlicensed}
\copyrightyear{2026}
\acmYear{2026}
\setcopyright{cc}
\setcctype{by}
\acmConference[WWW '26]{Proceedings of the ACM Web Conference 2026}{April 13--17, 2026}{Dubai, United Arab Emirates}
\acmBooktitle{Proceedings of the ACM Web Conference 2026 (WWW '26), April 13--17, 2026, Dubai, United Arab Emirates}
\acmPrice{}
\acmDOI{**********}
\acmISBN{*************}

\begin{document}

\title{PAMAS: Self-Adaptive Multi-Agent System with Perspective Aggregation for Misinformation Detection}

\author{Zongwei Wang}
\email{zongwei@cqu.edu.cn}
\affiliation{%
  \institution{Chongqing University}
  \city{Chongqing}
  \country{China}
}

\author{Min Gao}
\email{gaomin@cqu.edu.cn}
\authornote{Corresponding author. The code is available at \url{https://github.com/CoderWZW/PAMAS}}
\affiliation{%
  \institution{Chongqing University}
  \city{Chongqing}
  \country{China}
}
\additionalaffiliation{%
  \institution{Key Laboratory of Dependable Service Computing in Cyber Physical Society (Chongqing University), Ministry of Education of China}%
}

\author{Junliang Yu}
\email{junl.yu@outlook.com}
\affiliation{%
  \institution{Griffith University}
  \city{Brisbane}
  \country{Australia}}

\author{Tong Chen}
\email{tong.chen@uq.edu.au}
\affiliation{%
  \institution{The University of Queensland}
  \city{Brisbane}
  \country{Australia}}

\author{Chenghua Lin}
\email{chenghua.lin@manchester.ac.uk}
\affiliation{%
  \institution{University of Manchester}
  \city{Manchester}
  \country{United Kingdom}}

\renewcommand{\shortauthors}{Zongwei Wang, Min Gao, Junliang Yu, Tong Chen \& Chenghua Lin}

\begin{abstract}
Misinformation on social media poses a critical threat to information credibility, as its diverse and context-dependent nature complicates detection. Large language model–empowered multi-agent systems (MAS) present a promising paradigm that enables cooperative reasoning and collective intelligence to combat this threat. However, conventional MAS suffer from an information-drowning problem, where abundant truthful content overwhelms sparse and weak deceptive cues. With full input access, agents tend to focus on dominant patterns, and inter-agent communication further amplifies this bias. To tackle this issue, we propose PAMAS, a multi-agent framework with perspective aggregation, which employs hierarchical, perspective-aware aggregation to highlight anomaly cues and alleviate information drowning. PAMAS organizes agents into three roles: Auditors, Coordinators, and a Decision-Maker. Auditors capture anomaly cues from specialized feature subsets; Coordinators aggregate their perspectives to enhance coverage while maintaining diversity; and the Decision-Maker, equipped with evolving memory and full contextual access, synthesizes all subordinate insights to produce the final judgment. Furthermore, to improve the efficiency in multi-agent collaboration, PAMAS incorporates self-adaptive mechanisms for dynamic topology optimization and routing-based inference, enhancing both efficiency and scalability. Extensive experiments on multiple benchmark datasets demonstrate that PAMAS achieves superior accuracy and efficiency, offering a scalable and trustworthy way for misinformation detection. 


\end{abstract}




\maketitle

\section{Introduction}
Misinformation on social media has emerged as a serious threat to information credibility. It manifests in diverse forms, such as fake user accounts~\cite{42wang2025id}, deceptive product reviews~\cite{26yang2021rumor}, and manipulated news~\cite{25shu2019defend}, all of which distort public perceptions and erode trust in digital ecosystems. To contain these risks, in the early stages, platforms primarily relied on human auditors to identify and remove harmful content. Yet as platforms expand and the volume of user-generated content grows exponentially, the capacity of human auditors alone is quickly outpaced~\cite{28fahfouh2022contextual}. Deep learning models have therefore been adopted to assist auditors, but their opaque decisions often lack interpretability and fail to provide credible explanations~\cite{30sharma2022detecting,54wang2023efficient}. As a result, auditors still need to re-examine many cases to ensure accountability, which further increases their workload, leaving large amounts of harmful content unchecked.

\begin{figure}[t]
    \centering
    \subfloat[Existing MAS suffer from the information-drowning problem. The agents marked by red dashed boxes may initially detect anomalies, but their signals are later overwhelmed and disappear during aggregation.]{%
        \includegraphics[width=0.48\textwidth]{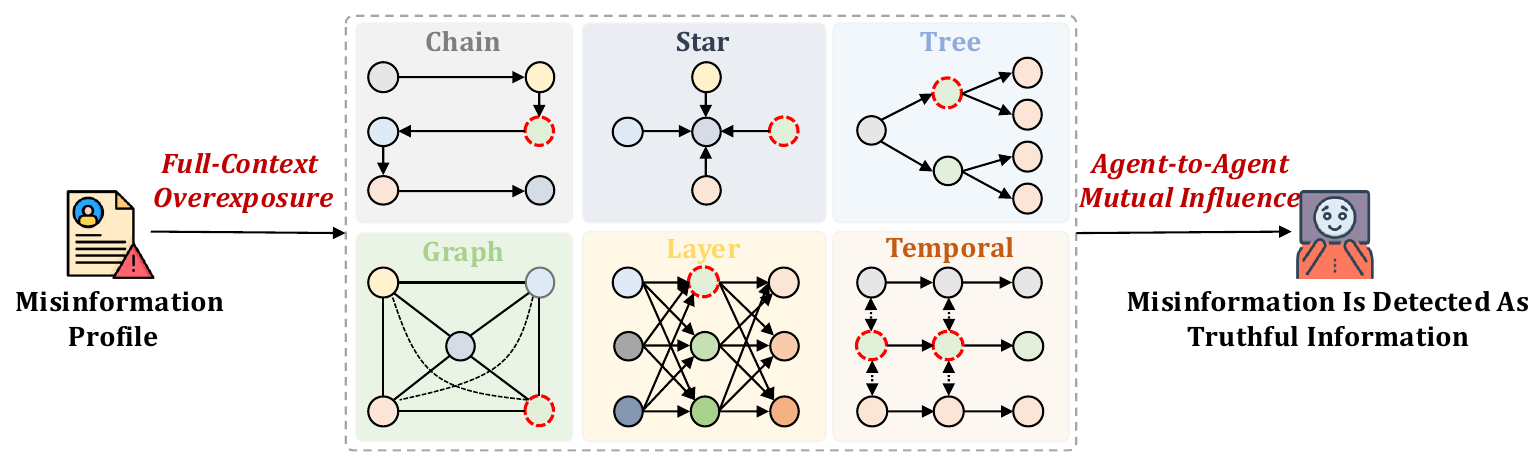}%
        \label{fig:image1}
    }
    \hfill
    \subfloat[ Multi-Agent System with Perspective Aggregation (PAMAS). The agents marked by red dashed boxes detect anomaly cues that are preserved and transmitted to the Decision-Maker for the final correct judgment.]{%
        \includegraphics[width=0.48\textwidth]{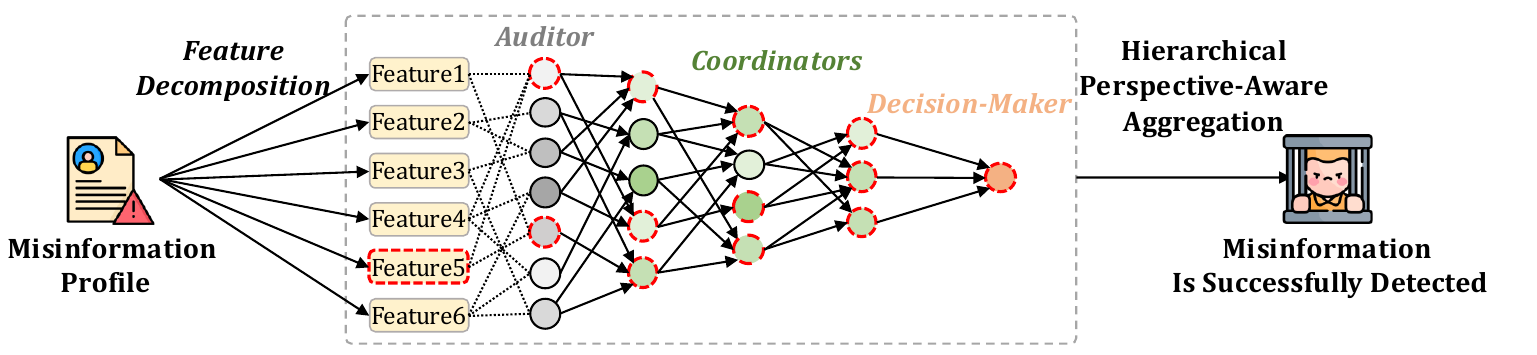}%
        \label{fig:image2}
    }
    \caption{Comparison between the existing MAS design and our proposed PAMAS.}
    \label{introduction}
\end{figure}

Given these limitations, large language models (LLMs) provide new avenues for improving content moderation~\cite{31liang2023encouraging,32wang2024mixture}. Unlike conventional deep learning models, LLMs exhibit strong reasoning and contextual understanding capabilities, enabling more transparent and interpretable decisions~\cite{34cobbe2021training,35chen2021evaluating}.
 Building upon these strengths, recent research shows that multi-agent systems (MAS) can outperform single agents on complex tasks by fostering cooperative reasoning across diverse collaboration topologies ~\cite{37tran2025multi}, such as chain~\cite{18qian2024chatdev}, star~\cite{20wu2024autogen}, tree~\cite{21ishibashi2024self}, graph~\cite{06zhang2025g}, layer~\cite{23zhuge2024gptswarm}, and temporal structure~\cite{07liu2024dynamic}. These advances naturally motivate MAS frameworks that collaborate with human auditors, where LLM-based agents not only generate automated judgments but also offer interpretable reasoning to support human oversight.


However, conventional MAS suffer from an information-drowning problem in misinformation detection, where abundant truthful content overwhelms sparse, fragmented and weak deceptive signals. As illustrated in Fig.~\ref{fig:image1}, existing MAS typically expose each agent to the complete input context of the target information to ensure comprehensive reasoning, yet this approach inadvertently amplifies the dominance of normal signals, causing agents to converge toward overly benign decisions. The problem is further intensified during agent-to-agent communication and debates, where correlated judgments reinforce one another, suppressing minority anomaly evidence and leaving misinformation even more deeply concealed.

To overcome the information-drowning problem, we design PAMAS, a novel MAS framework with Perspective Aggregation for misinformation detection, which employs hierarchical perspective-aware aggregation to actively highlight anomaly cues and mitigate information drowning. As shown in Fig.~\ref{fig:image2}, the information is first decomposed in a perspective-aware manner, where different feature subsets are extracted to emphasize distinct aspects of the content. Building on these specialized perspectives, PAMAS progressively aggregates the resulting judgments through a hierarchical structure of three roles, including Auditors, Coordinators, and a Decision-Maker. Auditors serve as the first layer of observation, each specializing in a specific feature subset to capture amplified anomalies. Their independent perspectives ensure diversity in detection and prevent the dominance of any single interpretation. Coordinators operate at the intermediate level, each responsible for supervising a subset of Auditors and integrating their reports into more comprehensive judgments. Through localized aggregation, they broaden the analytical scope while preserving the independence of underlying Auditor outputs, ensuring that amplified anomaly cues are retained through intermediate reasoning. At the top of the hierarchy, the Decision-Maker synthesizes its own direct judgment with the refined insights from all Coordinators. Equipped with self-evolving memory and full contextual access, it continuously enhances its reasoning capability, reconciles conflicting views across subordinates, preserves minority yet high-risk signals, and delivers the final decision.

Although this hierarchical perspective-aware aggregation enhances robustness in misinformation detection, operating PAMAS under conventional optimization and inference mechanisms would be both time-consuming and token-intensive. On the one hand, within each layer, every agent of the same role independently performs judgments, leading to repeated LLM calls and redundant token consumption. On the other hand, across layers, complex inter-agent interactions require extensive exchanges of inputs and outputs, further increasing inference time and token usage. To improve efficiency, we enhance PAMAS with self-adaptive mechanisms at both the optimization and inference stages, enabling the system to dynamically adjust its topology and refine agent capabilities over training. In the optimization stage, a topology adaptation strategy prunes redundant or overly similar agents to reduce wasted computation, while introducing new agents when existing structures exhibit correlated failures; subsequently, a targeted correction strategy enables misclassification-driven self-improvement, where only agents responsible for erroneous judgments are updated: Auditors refine their memory to improve localized experience, Coordinators adjust trust weights over their supervised subordinates, and the Decision-Maker consolidates these refinements into higher-level experiential guidance to prevent repeated mistakes. In the inference stage, a confidence-guided routing strategy selectively activates the most reliable branches instead of traversing the entire hierarchy, thereby substantially lowering computational overhead while preserving decision quality.

The contributions of this paper are four-fold:
\begin{itemize}[leftmargin=*]
    \item To the best of our knowledge, we are the first to systematically incorporate LLM-empowered MAS into misinformation detection and reveal the inherent information-drowning problem within this setting.
    \item We propose PAMAS, which performs hierarchical perspective-aware aggregation to amplify anomaly cues and consolidate heterogeneous judgments. 
    \item We design efficiency-oriented self-adaptive strategies for PAMAS, including dynamic topology adaptation, targeted agent refinement, and confidence-guided routing, which jointly improve the effectiveness and scalability.  
    \item We extensively validate our method across multiple benchmark datasets covering diverse misinformation scenarios, including abnormal users, misleading reviews, and fake news, where our method consistently outperforms strong baselines in terms of accuracy, efficiency, and interpretability.
\end{itemize}

\section{Preliminaries}
We begin by formalizing the setting of a single agent for misinformation detection.  
Each agent $a$ is equipped with three essential components:  
a \textit{profile} $\mathcal{P}_a$ that defines its persona and perspective for making judgments,  
a \textit{memory} $\mathcal{M}_a$ that records its historical experiences and past decisions,  
and a reasoning function $\phi_a$ that performs inference.  

Given an input instance $x \in \mathcal{X}$, agent $a$ produces a binary judgment and an associated explanation:
\begin{equation}
(d_a, r_a) \sim \phi_a(x, \mathcal{P}_a, \mathcal{M}_a),
\end{equation}
where $d_a \in \{0,1\}$ indicates whether $x$ is classified as misinformation, and $r_a \in \mathcal{R}$ is the natural-language explanation.  

Then, we extend to a multi-agent system $\mathcal{A} = \{a_1, a_2, \dots, a_m\}$, in which agents operate with heterogeneous profiles and distinct memories.  
Importantly, agents are not isolated: there exist relations among them such that the output of certain agents becomes part of the input to others.  

Formally, if $a_j$ is related to a set of agents $\mathcal{N}(j) \subseteq \mathcal{A}$,  
then $a_j$ takes into account both the external input $x$ and the results of all related agents when performing inference:
\begin{equation}
(d_{a_j}, r_{a_j}) \sim \phi_{a_j}\!\left(x, \{(d_{a_k}, r_{a_k}) \mid a_k \in \mathcal{N}(j)\}, \mathcal{P}_{a_j}, \mathcal{M}_{a_j}\right).
\end{equation}

At the system level, the collective inference function $\Phi$ aggregates the outputs of multiple agents:
\begin{equation}
(D, R) \sim \Phi\big(\{(d_{a_j}, r_{a_j})\}_{j=1}^m\big),
\end{equation}
yielding the decision $D \in \{0,1\}$ and a synthesized explanation $R$.  

\section{PAMAS: Self-Adaptive Multi-Agent System with Perspective Aggregation}
In this section, we introduce the PAMAS, which hierarchically organizes agents and aggregates their perspectives through a self-adaptive mechanism, thereby enhancing robustness and efficiency in misinformation detection. We first describe the distinct responsibilities of these roles (Sec.~3.1), followed by the system initialization and topology construction (Sec.~3.2), and the proposed self-adaptive strategies applied during optimization (Sec.~3.3) and inference (Sec.~3.4). The corresponding prompt designs used in PAMAS are provided in Appendix~\ref{prompts}.

\begin{figure*}[t]
  \centering
  \includegraphics[width=1\linewidth]{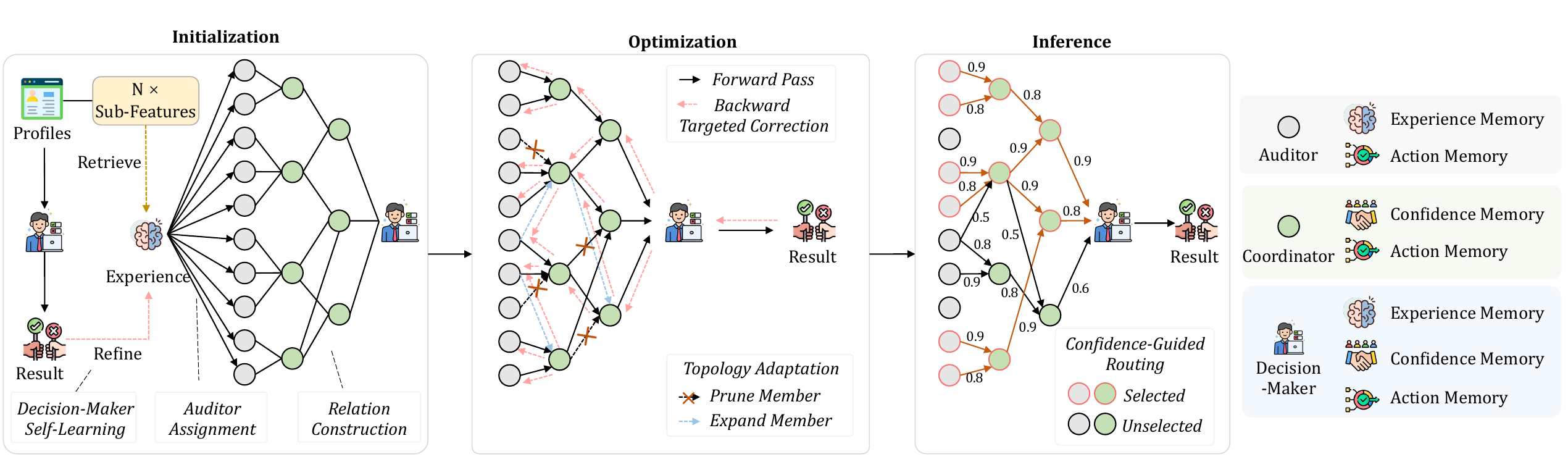} 
  \caption{The framework of PAMAS, including the process of initialization, optimization, and inference.}
\label{framework}
\end{figure*}

\subsection{Agent Roles}
PAMAS organizes agents into clear relations across three roles: \textit{Auditors}, \textit{Coordinators}, and a \textit{Decision-Maker}.
\subsubsection{Auditor}  
Auditors are the fundamental detection units, and also serve as the basic building block of the input layer in PAMAS. To ensure diversity of perspectives, we extract multiple dimensions of characteristics from each instance to be examined (e.g., statistical ratings, textual semantics such as sentiment polarity or consistency across statements, and temporal dynamics). Each Auditor $A$ is assigned a distinct \textit{profile} $\mathcal{P}_A$, which restricts its observation to a particular subset of characteristics rather than the full set of dimensions. By doing so, different Auditors develop complementary profiles, each focusing on specific aspects of the target detected information. More details on the characteristic dimensions used for profiling Auditors are provided in Appendix~\ref{Characteristic}.

The memory of an Auditor $A$ is divided into two components:  

- \textit{Experience memory} $\mathcal{M}^{\text{exp}}_A$, which summarizes detection experience distilled from accumulated cases. This forms the agent’s evolving knowledge base for identifying misinformation.  

- \textit{Action memory} $\mathcal{M}^{\text{act}}_A$, which logs the concrete decisions the agent has made on past inputs, together with the reasoning statements that supported them. This enables backtracking of how specific judgments were reached.    

\begin{tcolorbox}[
  enhanced,            
  colback=gray!10,
  colframe=gray,
  width=\linewidth,
  arc=1mm,
  auto outer arc,
  title={Auditor Memories},
  breakable,
  boxsep=0mm,
  top=2mm,    
  bottom=2mm, 
  fontupper=\linespread{0.9}\selectfont
]
\textbf{Experience memory sample:}  
``Misinformation accounts often reuse identical textual templates across items while maintaining abnormally high sentiment intensity.''  

\textbf{Action memory sample:}  
\{\texttt{user\_id}: 1327, \texttt{decision}: 1, \texttt{reason}: ``reviews are excessively repetitive and sentimentally exaggerated''\}  
\end{tcolorbox}

Given an input $x$, the Auditor produces a decision–reason pair:
\begin{equation}
(d_A, r_A) \sim \phi_A(x , \mathcal{P}_A, \mathcal{M}^{\text{act}}_A, \mathcal{M}^{\text{exp}}_A),
\label{reason_auditor}
\end{equation}
where $d_A \in \{0,1\}$ is the predicted label, and $r_A \in \mathcal{R}$ is the explanation.  
The output $(d_A, r_A)$ is simultaneously stored in action memory for accountability.  

\subsubsection{Coordinator}  
A Coordinator acts as the intermediate organizer, considering outputs from subordinate Auditors or lower-level Coordinators, and structurally corresponds to the hidden layer of PAMAS. While Auditors are responsible for making first-order judgments from restricted feature subsets, Coordinators focus on aggregating these heterogeneous perspectives into more reliable decisions. This hierarchical relationship ensures that anomaly cues identified by individual Auditors are properly amplified, and consolidated before being passed upward for higher-level reasoning.

The memory of a Coordinator $C$ is divided into two components:  

- \textit{Confidence memory} $\mathcal{M}^{\text{conf}}_C$, which records the trust weights $w_j$ assigned to each subordinate. These weights evolve with optimization, reflecting the reliability of different subordinates over time.  

- \textit{Action memory} $\mathcal{M}^{\text{act}}_C$, which logs the aggregated decisions made by $C$, along with the subordinate votes that contributed to them.

\begin{tcolorbox}[
  enhanced,            
  colback=gray!10,
  colframe=gray,
  width=\linewidth,
  arc=1mm,
  auto outer arc,
  title={Coordinator Memories},
  breakable,
  boxsep=0mm,
  top=2mm,    
  bottom=2mm, 
  fontupper=\linespread{0.9}\selectfont
]
\textbf{Confidence memory sample:}  
\{\texttt{subordinate\_id}: 2, \texttt{weight}: 0.73; \texttt{subordinate\_id}: 15, \texttt{weight}: 0.41 \}  

\textbf{Action memory sample:}  
\{ \texttt{decision}: 1, \texttt{votes}: [ (1,0.73), (0,0.41) ] \}  

\end{tcolorbox}

Given subordinate outputs $\{(d_j, r_j)\}$, the Coordinator aggregates them by weighted voting:
\begin{equation}
d_C = \mathbb{I}\!\left[\sum_{j \in \mathcal{S}(C)} w_j \cdot (2d_j-1) > 0\right],
\label{reason_leadership}
\end{equation}
where $\mathcal{S}(C)$ is the subordinates of Coordinator $C$, and stores both the decision and vote distribution into its action memory.  

Coordinators can themselves be recursively organized into multiple layers, whose depth can be flexibly adjusted to fit task complexity, forming a scalable hierarchy. At the top of this hierarchy, the highest-level Coordinator interacts directly with the Decision-Maker, which synthesizes the final decision.

\subsubsection{Decision-Maker}  
At the top of the hierarchy, the Decision-Maker serves as the final arbiter, synthesizing its own direct judgment $\phi_D(x)$ with the aggregated outputs of all top-level Coordinators. It maintains three memories: \textit{confidence memory} $\mathcal{M}^{\text{conf}}_D$ toward Coordinators, \textit{experience memory} $\mathcal{M}^{\text{exp}}_D$ distilled from repeated interactions, and \textit{action memory} $\mathcal{M}^{\text{act}}_D$ archiving final decisions and supporting reasons. Its inference rule is defined as:
\begin{equation}
(d_D, r_D) \sim \phi_D\!\left(x , \{(d_{C_k}, r_{C_k}, \alpha_k)\}_{k}, \mathcal{P}_D, 
\mathcal{M}^{\text{conf}}_D, \mathcal{M}^{\text{exp}}_D, \mathcal{M}^{\text{act}}_D \right),
\label{reason_boss}
\end{equation}
where ${(d_{C_k}, r_{C_k}, \alpha_k)}$ are the outputs and trust weights of top-level Coordinators. Since a Coordinator does not independently generate reasoning, $r_{C_k}$ is inherited from its subordinates and taken as the explanation provided by the highest-weight Auditor whose decision aligns with the Coordinator’s aggregated output. $\phi_D$ combines them with the Decision-Maker’s profile $\mathcal{P}_D$ and memories to yield the final decision $d_D$ and explanation $r_D$. Here, $\mathcal{P}_D$ specifies the task of the Decision-Maker integrating all highest-level Coordinator outputs while leveraging the full set of dimensions and accumulated memories to derive the final judgment.

\subsection{Initialization of PAMAS}
Although the three roles are defined, how to initialize these agents and how to determine the relations between agents remain unresolved. This section describes the initialization of PAMAS. At first, all memories are created in empty form, and all trust weights (e.g., confidence weights $\alpha_k$ between superiors and subordinates) are initialized to $1$ to provide an unbiased starting point. Then, the process begins with the Decision-Maker, which first undergoes self-learning to establish an initial experience memory. Building on this foundation, Auditors are assigned profile-specific feature subsets and initialize their own experience through Decision-Maker-guided retrieval and self-reflection. Finally, relations between agents are established via hierarchical clustering, where each upper-level agent selects a subset of lower-level agents to form localized groups, defining the multi-layer structure.

\subsubsection{Decision-Maker Self-Learning.}  We first initialize the system by training the Decision-Maker through a self-learning process. 
Given an input instance $x$ with ground-truth label $y$, the Decision-Maker forms a preliminary judgment $(\hat{d}_D,\hat{r}_D)$ based on Equation (\ref{reason_boss}) and then performs self-reflection by contrasting it with $y$ to distill new experience:
\begin{equation}
\mathcal{M}^{\text{exp}}_D \leftarrow \mathcal{M}^{\text{exp}}_D \cup \operatorname{SelfRef}_D\!\big(x, \hat{d}_D, \hat{r}_D, y\big),
\end{equation}
where $\operatorname{SelfRef}_D(\cdot)$ summarizes generalized experience from the input $(x,\hat{d}_D,\hat{r}_D,y)$ and appends them to the experience memory. 
This establishes the initial experience memory of the Decision-Maker, forming a knowledge base for subsequent training of Auditors.

\subsubsection{Auditor Assignment.} To transfer knowledge effectively while avoiding homogeneous behaviors, each input is decomposed into multiple feature dimensions, and every Auditor is assigned a distinct profile $\mathcal{P}_a$ that restricts its observation to a subset of these dimensions. This ensures that Auditors specialize in complementary perspectives rather than sharing identical views.

However, feature allocation alone is coarse, since all agents would otherwise start with identical experience memories, undermining their diversity. To address this, we introduce a \textit{Decision-Maker-Guided Auditor Training}, in which the Decision-Maker provides experiential knowledge to each Auditor. Inspired by retrieval-augmented generation (RAG)~\cite{40lewis2020retrieval}, Auditors learn from the Decision-Maker’s experience memory by retrieving fragments most relevant to their profiles.

Formally, given Auditor $A$’s observable feature subset $\mathcal{F}(x)_A$, we encode its textual description into an embedding vector $\mathbf{z}_A$. Each experience fragment $\mathbf{e} \in \mathcal{M}^{\text{exp}}_D$ is represented both by its raw description and corresponding embedding representation. Auditor $A$ then retrieves the top-$k$ relevant fragments:
\begin{equation}
E_A \;=\; \operatorname{TopK}_k\!\Big(\,\mathrm{Sim}(\mathbf{z}_A,\mathbf{e})\Big),
\label{expupdate1}
\end{equation}
where $\text{Sim}(\cdot,\cdot)$ denotes cosine similarity.

The textual content corresponding to retrieved fragments $E_A$ is distilled by Auditor $A$ through a local self-reflection process:
\begin{equation}
\mathcal{M}^{\text{exp}}_A \leftarrow \operatorname{SelfRef}_A(E_A),
\label{expupdate2}
\end{equation}
where $\operatorname{SelfRef}_A(\cdot)$ summarizes the fragments into generalizable heuristics. This completes the initialization of each Auditor’s profile and experience memory.

\subsubsection{Relation Construction.}  
After Auditors are assigned, we establish the relations among the three roles of agents, grouping those with similar behaviors into organized teams and thereby defining the structure. Specifically, Auditors are first clustered ~\cite{41murtagh2012algorithms} based on the similarity of their label prediction vectors on the validation set. For each cluster $\mathcal{G}$, one Coordinator $C$ is designated to manage it. The Coordinator’s anchor Auditor is selected as the member achieving the highest validation F1-score:
\begin{equation}
A^* = \arg\max_{A \in \mathcal{G}} \text{F1}(A).
\end{equation}
After the anchor is determined, the corresponding group for this Coordinator is expanded by randomly sampling additional Auditors from $\mathcal{G}$ until the predefined group size is reached. If the number of available Auditors in $\mathcal{G}$ is insufficient, the nearest neighboring cluster is used for supplementary sampling.
Each resulting group thus defines the subordinates of a Coordinator:
\begin{equation}
\mathcal{S}(C) = \{A^*, A_2, \dots, A_k\}, \quad k \leq n_{\max},
\end{equation}
where $n_{\max}$ is the maximum allowable group size set for each coordinator. This procedure is recursively applied: newly formed Coordinators are clustered in the same manner, anchors are selected, and higher-level groups are built until the Decision-Maker sits at the top.

As a result, PAMAS forms a hierarchical structure that embeds diversity, where each Auditor focuses on a distinct feature-based perspective and each Coordinator supervises a unique subset of members, jointly ensuring heterogeneous viewpoints across the structure.

\subsection{The Optimization of PAMAS}
Once initialized, PAMAS operates an iterative refinement process, which consists of three connected stages.
The \textit{forward pass} first generates decisions in a bottom-up manner.  
Based on these results, \textit{topology adaptation} adjusts the hierarchy by pruning redundant agents and expanding new ones when failures occur.  
Finally, under the adapted structure, \textit{backward targeted correction} updates agent memories for misclassified cases.

\subsubsection{Forward Pass.}  
Given an input instance $x$, each Auditor generates a decision–reason pair following Eq.~\eqref{reason_auditor}, leveraging both its action and experience memories.  
Coordinators then aggregate the outputs of their subordinates by weighted voting as defined in Eq.~\eqref{reason_leadership}, using trust weights stored in their confidence memories.  
Finally, the Decision-Maker synthesizes the top-level decision and explanation according to Eq.~\eqref{reason_boss}, integrating both its own profile and the aggregated results from top-level leaders. All decisions produced at each level are simultaneously logged into the corresponding agents’ action memories, ensuring accountability and enabling subsequent refinement.

\subsubsection{Topology Adaptation.}  
After each batch of forward pass, the topology of PAMAS adapts to balance efficiency and robustness through two complementary operations, pruning redundant members and expanding complementary members, both conducted within the scope of each Coordinator’s subordinate group.  

In pruning redundant members, each subordinate $i \in \mathcal{S}(C)$ of Coordinator $C$ is evaluated by a score:
\begin{align}
\text{Score}_i &= 
\underbrace{\big[\text{Acc}(\mathcal{S}(C)) - \text{Acc}(\mathcal{S}(C)\setminus \{i\})\big]}_{\text{marginal accuracy improvement}} \notag \\
&\quad - \lambda \cdot 
\underbrace{\frac{1}{|\mathcal{S}(C)|-1}\sum_{j \in \mathcal{S}(C)\setminus \{i\}} \cos(\mathbf{v}_i,\mathbf{v}_j)}_{\text{average redundancy with peers}},
\end{align}
where $\mathbf{v}_i$ denotes the prediction vector of agent $i$ on the validation set. $\lambda$ controls the balance of accuracy and redundancy. If  $\text{Score}_i < 0$, the agent is considered unhelpful and removed.  

In expanding complementary members, when a group $\mathcal{S}(C)$ exhibits correlated failures on borderline samples, we evaluate each candidate $i$ drawn from the lower layer but not belonging to $\mathcal{S}(C)$:
\begin{equation}
\text{Gain}_i = 
\underbrace{\Delta \text{Acc}(\mathcal{S}(C)\cup\{i\})}_{\text{expected accuracy improvement}} 
- \gamma \cdot 
\underbrace{\frac{1}{|\mathcal{S}(C)|}\sum_{j \in \mathcal{S}(C)} \cos(\mathbf{v}_i,\mathbf{v}_j)}_{\text{similarity penalty}},
\end{equation}
where $\Delta \text{Acc}(\mathcal{S}(C)\cup\{i\})$ denotes the improvement after adding $i$, and the second term penalizes overlap with existing members based on their prediction vectors.  
Candidates with the highest gain are incorporated as new subordinates under Coordinator $C$.

\subsubsection{Backward Targeted Correction.}  
Following topology adaptation, agents update their memories under the guidance of ground-truth labels, and this update is applied only to those agents that made incorrect judgments. 
This process proceeds in a top-down manner, from the Decision-Maker to Coordinators and finally to Auditors.  
The Decision-Maker first performs self-reflection, updating both its experience memory and its confidence in subordinates. For updating the experience memory:
\begin{equation}
\mathcal{M}^{\text{exp}}_D \leftarrow \mathcal{M}^{\text{exp}}_D \cup \operatorname{SelfRef}_D(x,\hat{d}_D,\hat{r}_D,y).
\end{equation}

For confidence updates, Decision-Maker updates its confidence memory for each subordinate $j$:
\begin{align} &\mathcal{M}^{\text{conf}}[j] \leftarrow (1-\alpha)\cdot \mathcal{M}^{\text{conf}}[j] \notag \\ &+ \alpha \cdot \Big[ \underbrace{f_{\text{ref}}(y,d_j)}_{\text{self-reflection term}} + \underbrace{\tfrac{\#\{d_j = y\}}{\#\text{samples}}}_{\text{historical accuracy term}} \Big], \end{align}
where the first term $f_{\text{ref}}(y,d_j)$ captures the outcome of the agent’s self-reflection process, i.e., the value extracted from its reasoning that reflects how the prediction–ground truth discrepancy influences subsequent weight adjustment.
The second term represents the long-term reliability of agent $j$, quantified as its historical accuracy on validation data. 
The $\alpha$ controls the balance between short-term reflective adjustment and long-term statistical evidence.

The same refinement logic is subsequently applied to Coordinators: each Coordinator updates its confidence memory in its own subordinates following the above procedure.
  
Auditors, as the most numerous and token-intensive role, do not perform costly self-reflection for each error.  
Instead, at the end of each epoch, Auditors update their experience memory by retrieving new fragments from the Decision-Maker’s experience memory, guided by the fragments already stored in their own memory. Specifically, Auditor $A$ uses its existing fragments to query $\mathcal{M}^{\text{exp}}_D$, retrieves the top-$k$ relevant fragments $E_A$ according to cosine similarity, and then applies the same self-reflection process to summarize them:
\begin{equation}
\mathcal{M}^{\text{exp}}_A \leftarrow \mathcal{M}^{\text{exp}}_A \cup \operatorname{SelfRef}_A(E_A).
\end{equation}
This incremental update enables Auditors to continually enrich their experience with Decision-Maker knowledge while avoiding costly per-instance reflection.

\subsection{Routing Inference of PAMAS}
After the optimization of PAMAS, the system can be deployed for inference. A straightforward deployment would require every agent to evaluate every incoming instance, which introduces substantial unnecessary computation: many simple instances can already be resolved with high agreement among lower-level agents, making further evaluations by higher-level agents redundant. To avoid such inefficiency, PAMAS employs a top-down, confidence-guided routing strategy, enabling the system to terminate inference early when subordinate consensus is sufficiently strong and to escalate only those contentious instances that require higher-level judgment.

Suppose the Decision-Maker $D$ supervises subordinates $\mathcal{S}(D)$ with confidences $\alpha_j$ recorded in $\mathcal{M}^{\text{conf}}_D$. Routing begins by activating the top-2 subordinates:
\begin{equation}
\{j_1,j_2\} = \arg\max\nolimits^{(2)}_{j \in \mathcal{S}(D)} \alpha_j,
\end{equation}
where $\arg\max^{(2)}$ returns the two indices with the highest confidences.

If their predictions agree, finalize $d_C = d_{j_1}$. Otherwise, activate more subordinates in descending confidence. At step $t$, the active set is:
\begin{equation}
\mathcal{A}_t = \{ j_1, j_2, j_3, j_4, \dots, j_{2t+1}, j_{2t+2} \}.
\end{equation}

Define the majority margin over the active set:
\begin{equation}
m(\mathcal{A}_t) = \sum_{j \in \mathcal{A}_t} (2d_j - 1),
\end{equation}
where we early stop as soon as a non-tie majority emerges, i.e., $m(\mathcal{A}_t) \neq 0$. Denote the minimal active set achieving this by $\mathcal{A}^{\star}$, with majority decision:
\begin{equation}
d_C = \mathbb{I}[\, m(\mathcal{A}^{\star}) > 0 \,].
\end{equation}

The same routing recursively proceeds top-down for all Coordinators, propagating until the lowest-level Auditors.
At this stage, the Decision-Maker itself constructs its own minimal majority-active set $\mathcal{A}^\star$. 
Each Coordinator selected into $\mathcal{A}^\star$ produces a routed decision $d_{C_k}$, which is then weighted by the Decision-Maker’s confidence $\alpha_k \in \mathcal{M}^{\text{conf}}_D$. 
Finally, the Decision-Maker integrates these routed summaries together with its own experience memory and the full feature set $\mathcal{F}(x)$ to produce the final decision and explanation:
\begin{equation}
(d_D, r_D) \sim \phi_{D}\!\Big(\mathcal{F}(x), \;
\{(d_{C_k}, r_{C_k}, \alpha_k)\}_{k \in \mathcal{A}^\star},\;
\mathcal{M}^{\text{conf}}_D, \;\mathcal{M}^{\text{exp}}_D, \;\mathcal{M}^{\text{act}}_D
\Big).
\end{equation}

A complete view of the algorithmic workflow is provided in the Appendix, referred to Algorithm~\ref{alg:hpamas}, and further analysis of PAMAS about interpretability and efficiency can be found in Appendix~\ref{analysis}.

\section{Experiments}
This section evaluates the effectiveness of PAMAS.  
Specifically, we aim to answer the following questions:  
(Q1) How does PAMAS perform compared to existing deep learning methods, as well as MAS methods? 
(Q2) What is the impact of the proposed topology adaptation, targeted correction, and confidence-guided routing strategies on detection performance? 
(Q3) How efficient is PAMAS in terms of token consumption compared to other multi-agent methods? 
(Q4) How does PAMAS behave in case studies involving both correctly classified and misclassified instances? Experimental settings can be found in Appendix \ref{appendix:setting}.

\begin{table*}[t]
\caption{Performance comparison of PAMAS against deep learning and multi-agent baselines on three misinformation detection datasets. The best results are in bold, and the runners-up are underlined.}
\vspace{-10pt}
\centering
\resizebox{\textwidth}{!}{
\begin{tabular}{cccccccccccc|ccccc}
\hline
\multicolumn{2}{c|}{\multirow{2}{*}{\textbf{Method}}} &
  \multicolumn{5}{c|}{\textbf{Amazon}} &
  \multicolumn{5}{c|}{\textbf{DeRev2018}} &
  \multicolumn{5}{c}{\textbf{PolitiFact}} \\ \cline{3-17} 
\multicolumn{2}{c|}{} &
  \textbf{Accuracy} &
  \textbf{Precision} &
  \textbf{Recall} &
  \textbf{F1} &
  \multicolumn{1}{c|}{\textbf{AUC}} &
  \textbf{Accuracy} &
  \textbf{Precision} &
  \textbf{Recall} &
  \textbf{F1} &
  \textbf{AUC} &
  \textbf{Accuracy} &
  \textbf{Precision} &
  \textbf{Recall} &
  \textbf{F1} &
  \textbf{AUC} \\ \hline
\multicolumn{1}{c|}{\multirow{4}{*}{\begin{tabular}[c]{@{}c@{}}Deep Learning\\ Method\end{tabular}}} &
  \multicolumn{1}{c|}{MLP} &
  0.9235 &
  0.9023 &
  0.8823 &
  0.8913 &
  \multicolumn{1}{c|}{0.9224} &
  0.7947 &
  0.7825 &
  0.8135 &
  0.7945 &
  0.8524 &
  0.9148 &
  0.8900 &
  0.9106 &
  0.8994 &
  0.9136 \\
\multicolumn{1}{c|}{} &
  \multicolumn{1}{c|}{NFGCN-TIA} &
  0.9412 &
  0.9324 &
  0.9543 &
  0.9445 &
  \multicolumn{1}{c|}{0.9464} &
  \textbf{-} &
  - &
  - &
  - &
  - &
  - &
  - &
  - &
  - &
  - \\
\multicolumn{1}{c|}{} &
  \multicolumn{1}{c|}{SIPUL} &
  - &
  - &
  - &
  - &
  \multicolumn{1}{c|}{-} &
  0.9074 &
  0.8884 &
  0.9130 &
  0.8995 &
  0.9285 &
  - &
  - &
  - &
  - &
  - \\
\multicolumn{1}{c|}{} &
  \multicolumn{1}{c|}{BREAK} &
  - &
  - &
  - &
  - &
  \multicolumn{1}{c|}{-} &
  - &
  - &
  - &
  - &
  - &
  \underline{0.9465} &
  \underline{0.9478} &
  0.9493 &
  \underline{0.9502} &
  0.9425 \\ \hline
\multicolumn{1}{c|}{\multirow{8}{*}{\begin{tabular}[c]{@{}c@{}}Multi-Agent\\ Method\end{tabular}}} &
  \multicolumn{1}{c|}{Vanilla-LLM} &
  0.8956 &
  0.8765 &
  0.8856 &
  0.8788 &
  \multicolumn{1}{c|}{0.8832} &
  0.8157 &
  0.8066 &
  0.8226 &
  0.8182 &
  0.8281 &
  0.8723 &
  0.8529 &
  0.8751 &
  0.8637 &
  0.8765 \\
\multicolumn{1}{c|}{} &
  \multicolumn{1}{c|}{Vanilla-Agent} &
  0.9123 &
  0.9047 &
  0.8976 &
  0.8987 &
  \multicolumn{1}{c|}{0.9135} &
  0.8376 &
  0.8223 &
  0.8453 &
  0.8303 &
  0.8684 &
  0.9265 &
  0.8834 &
  0.9026 &
  0.8985 &
  0.9084 \\
\multicolumn{1}{c|}{} &
  \multicolumn{1}{c|}{Chain} &
  0.9151 &
  0.9234 &
  0.9025 &
  0.9124 &
  \multicolumn{1}{c|}{0.9584} &
  0.8463 &
  0.8535 &
  0.8849 &
  0.8685 &
  0.8735 &
  0.9343 &
  0.9238 &
  0.9478 &
  0.9383 &
  0.9305 \\
\multicolumn{1}{c|}{} &
  \multicolumn{1}{c|}{Star} &
  0.9135 &
  0.9337 &
  0.9186 &
  0.9234 &
  \multicolumn{1}{c|}{0.9653} &
  0.8764 &
  0.8834 &
  0.8949 &
  0.8839 &
  0.9036 &
  0.9378 &
  0.9432 &
  0.9385 &
  0.9436 &
  0.9427 \\
\multicolumn{1}{c|}{} &
  \multicolumn{1}{c|}{Tree} &
  0.9375 &
  \textbf{0.9484} &
  0.9274 &
  0.9355 &
  \multicolumn{1}{c|}{0.9535} &
  0.8925 &
  0.8885 &
  0.9054 &
  0.9025 &
  0.9134 &
  0.9314 &
  0.9282 &
  0.9402 &
  0.9393 &
  0.9332 \\
\multicolumn{1}{c|}{} &
  \multicolumn{1}{c|}{Graph} &
  0.9325 &
  0.9473 &
  0.9599 &
  0.9527 &
  \multicolumn{1}{c|}{{\ul 0.9696}} &
  0.8885 &
  {0.8985} &
  0.9085 &
  {0.9043} &
  0.9112 &
  0.9332 &
  {0.9456} &
  0.9325 &
  0.9374 &
  {\ul 0.9686} \\
\multicolumn{1}{c|}{} &
  \multicolumn{1}{c|}{Layer} &
  0.9347 &
  0.9385 &
  0.9464 &
  0.9389 &
  \multicolumn{1}{c|}{0.9683} &
  0.9255 &
  0.9082 &
  0.9174 &
  0.9112 &
  {0.9331} &
  0.9404 &
  0.9285 &
  0.9496 &
  0.9353 &
  0.9625 \\
\multicolumn{1}{c|}{} &
  \multicolumn{1}{c|}{DyLAN} &
  {\ul 0.9454} &
  0.9456 &
  {\ul 0.9696} &
  {\ul 0.9535} &
  \multicolumn{1}{c|}{0.9635} &
  {\ul 0.9314} &
  {\ul 0.9135} &
  {\ul 0.9328} &
  {\ul 0.9242}&
  {\ul 0.9425} &
  {0.9424} &
  0.9329 &
  {\ul 0.9523} &
  {0.9475} &
  0.9604 \\ \cline{2-17} 
\multicolumn{1}{c|}{} &
  \multicolumn{1}{c|}{PAMAS} &
  \textbf{0.9715} &
  \underline{0.9477} &
  \textbf{0.9852} &
  \textbf{0.9661} &
  \multicolumn{1}{c|}{\textbf{0.9735}} &
  \textbf{0.9628} &
  \textbf{0.9427} &
  \textbf{0.9739} &
  \textbf{0.9621} &
  \textbf{0.9648} &
  \textbf{0.9643} &
  \textbf{0.9553} &
  \textbf{0.9706} &
  \textbf{0.9594} &
  \textbf{0.9733} \\ \hline
\end{tabular}}
\label{performance comparison}
\end{table*}
  
\subsection{Performance Comparison}
We evaluate the proposed PAMAS against traditional and representative deep learning models, as well as state-of-the-art MAS methods. The results are reported in Table~\ref{performance comparison}, where we consider multiple evaluation metrics across three benchmark datasets. From the experimental results, we derive the following observations:
\begin{itemize}[leftmargin=*]
\item Compared with traditional deep learning methods and existing MAS baselines, PAMAS achieves the best performance across nearly all metrics and datasets, demonstrating the effectiveness of perspective-aware aggregation in misinformation detection.
\item The Vanilla setup achieves performance comparable to or even worse than the simple MLP baseline, and falls far behind multi-agent systems. This indicates that directly invoking an LLM for detection performs even worse. Although equipping it with domain-specific knowledge and evolved memory brings slight improvement, both LLM-based detection and relying on a single agent remain insufficient to handle the deceptive and context-dependent nature of misinformation.
\item Existing multi-agent topologies do not always outperform strong deep learning models. While simple multi-agent structures such as Chain, and Star achieve comparable or slightly better performance than MLP, they often underperform advanced deep learning methods tailored for specific tasks. Nevertheless, more sophisticated designs such as Graph, Layer, and DyLAN show performance improvements, highlighting that complex structural design is important in misinformation detection. 
\end{itemize}

\subsection{Ablation Study}
\begin{figure}[h]
  \centering
  \includegraphics[width=1\linewidth]{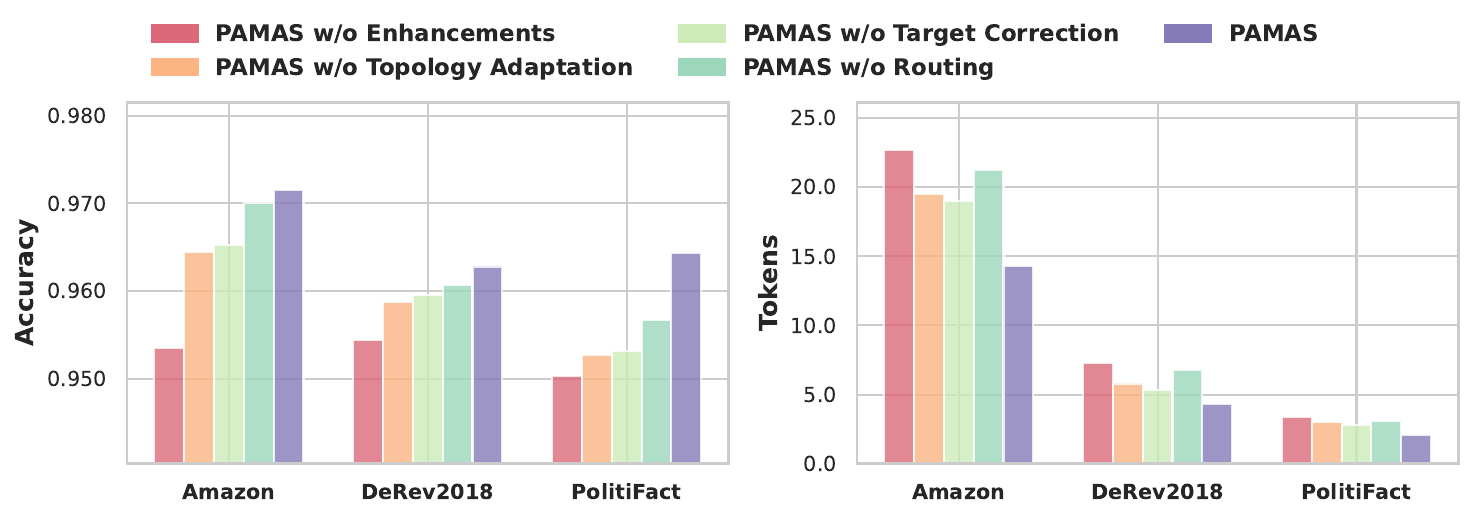} 
  \caption{Impact of topology adaptation, target correction, and confidence-guided routing on accuracy–efficiency trade-offs across three datasets.}
\label{ablation}
\end{figure}

We further evaluate the contribution of each proposed efficiency-oriented strategy, including topology adaptation, target correction, and confidence-guided routing. Specifically, we design five variants:  \textbf{PAMAS w/o Enhancements}, which disables all three strategies; \textbf{PAMAS w/o Topology Adaptation}, which retains target correction and routing but removes topology adaptation; \textbf{PAMAS w/o Target Correction}, which disables target correction while keeping the other two strategies; \textbf{PAMAS w/o Routing}, which removes confidence-guided routing but preserves the other strategies; and the \textbf{full PAMAS}, which integrates all three strategies. The results are summarized in Figure~\ref{ablation}, from which the following observations can be drawn:
\begin{itemize}[leftmargin=*]
\item Even without any enhancement, PAMAS achieves competitive performance while consuming fewer tokens than complex multi-agent structures such as Fully Graph, Fully Layer, and DyLAN. 
\item Each of the three strategies provides measurable improvements, both in terms of accuracy and token efficiency. Topology adaptation enhances robustness and accuracy by dynamically adjusting the number of active agents; target correction further strengthens detection by selectively retraining agents with a history of misjudgments, enabling precise correction of systematic weaknesses; and confidence-guided routing improves inference efficiency by focusing communication on the most reliable signals.
\item Among the three, topology adaptation delivers the largest accuracy gains across datasets, whereas target correction achieves the greatest reduction in token consumption. Confidence-guided routing adds a further efficiency boost with a modest but consistent accuracy lift, placing the full PAMAS on the best accuracy–efficiency frontier.
\end{itemize}

\subsection{Token Efficiency Comparison}
\begin{figure}[h]
  \centering
  \includegraphics[width=1\linewidth]{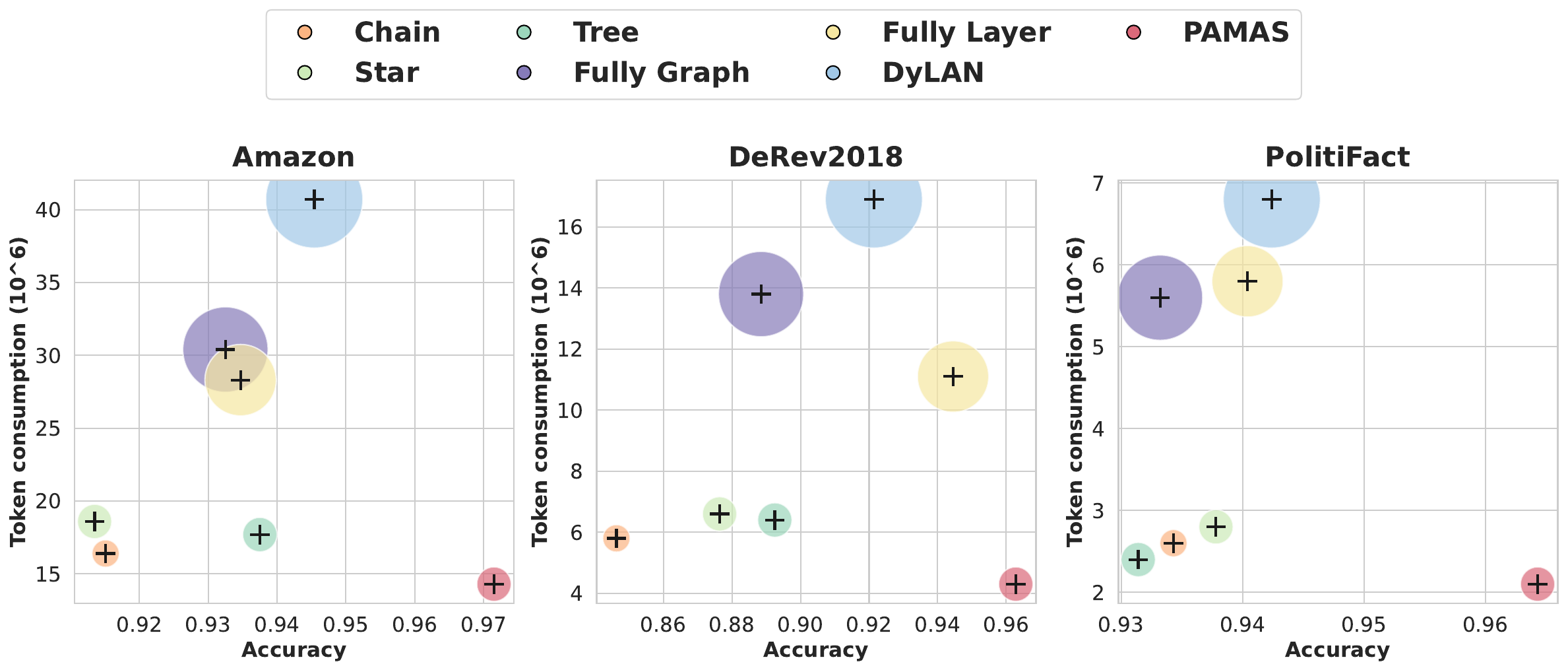} 
  \caption{Comparative results of various MAS in terms of accuracy and token consumption across three datasets.}
\label{token}
\end{figure}

We further compare PAMAS with existing multi-agent collaboration topologies in terms of token consumption, as shown in Figure~\ref{token}. From the results, two main findings emerge:

\begin{itemize}[leftmargin=*]
\item PAMAS achieves the lowest token consumption while maintaining the best performance. Due to its topology adaptation, target correction, and confidence-guided routing mechanisms, PAMAS avoids unnecessary updates and redundant activations, leading to a substantial reduction in tokens without sacrificing accuracy. This demonstrates its suitability for large-scale misinformation detection tasks where efficiency is critical.
\item Complex MAS generally incur higher token costs but yield better performance than simpler ones. Topologies such as fully connected graphs, layered organizations, or DyLAN exhibit stronger detection performance compared to basic chain or star structures, but at the expense of significantly higher token usage. While such designs may be reasonable for general-purpose reasoning tasks, they are inefficient for misinformation detection, where the combination of robustness and scalability is essential. PAMAS strikes a better balance by introducing structural hierarchy and adaptive mechanisms tailored to this domain.  
\end{itemize}

\subsection{Case Studies}
We present three representative cases to demonstrate how PAMAS performs: one successful case and two unsuccessful ones.

\begin{tcolorbox}[
enhanced, colback=green!5, colframe=green!20!black,
width=\linewidth, arc=1mm, auto outer arc,
title={Successful Case (Ground-truth: 1, Prediction: 1)}, breakable, boxsep=0mm,
top=2mm, bottom=2mm, fontupper=\linespread{0.95}\selectfont
]
\textbf{User23898, Decision: 1}  
\textbf{Reason:} \textcolor{blue}{All coordinators unanimously predict malicious (1) with high confidence}.  
The user's metrics, including \textcolor{blue}{abnormally high posting frequency}, \textcolor{blue}{overly extreme sentiment}, and \textcolor{blue}{repetitive suspicious content}, strongly support a malicious classification.  
There is \textcolor{blue}{no conflicting evidence} to suggest benign activity.
\end{tcolorbox}

\textbf{Analysis:}  
This case succeeds because both the coordinators and the decision-maker are fully aligned. The user’s features strongly indicate malicious behavior, and the absence of contradictory signals ensures a reliable prediction. 

\begin{tcolorbox}[
enhanced, colback=red!5, colframe=red!50!black,
width=\linewidth, arc=1mm, auto outer arc,
title={Unsuccessful Case 1: Coordinator Disagreement (Ground-truth: 0, Prediction: 1)}, breakable, boxsep=0mm,
top=2mm, bottom=2mm, fontupper=\linespread{0.95}\selectfont
]
\textbf{User18284, Decision: 1}  
\textbf{Reason:} \textcolor{red}{Votes are split}: 5 out of 8 coordinators predict malicious (1), while the rest predict benign (0).  
Malicious votes are driven by \textcolor{red}{inconsistent rating patterns and sentiment fluctuations}, whereas benign votes emphasize \textcolor{blue}{overall positive sentiment and stable review frequency}.  
\textcolor{red}{Conflicting evidence prevents a clear consensus}, leading the decision-maker to side with the majority malicious vote.
\end{tcolorbox}

\textbf{Analysis:}  
This case fails because the coordinators produced conflicting judgments. The decision-maker was forced to arbitrate between divided evidence, which reduced confidence and resulted in a misclassification.  

\begin{tcolorbox}[
enhanced, colback=red!5, colframe=red!50!black,
width=\linewidth, arc=1mm, auto outer arc,
title={Unsuccessful Case 2: Coordinator–Decision Mismatch (Ground-truth: 0, Prediction: 1)}, breakable, boxsep=0mm,
top=2mm, bottom=2mm, fontupper=\linespread{0.95}\selectfont
]
\textbf{User7430, Decision: 1}  
\textbf{Reason:} \textcolor{blue}{Most coordinators unanimously predict benign (0)}, citing \textcolor{blue}{high avg\_rating}, \textcolor{blue}{positive sentiment}, and \textcolor{blue}{consistent review length}.  
However, I override, highlighting \textcolor{red}{abnormal interaction frequency}, and \textcolor{red}{suspicious review}, which together suggest malicious behavior. 
\end{tcolorbox}

\textbf{Analysis:}  
This case fails because the decision-maker’s independent assessment contradicted the unanimous coordinator consensus. Although surface-level metrics suggested benign behavior, deeper anomaly signals led the decision-maker to predict malicious, creating a misalignment between consensus and final decision.  

\section{Related Work}
\noindent\textbf{Misinformation Detection.}
Misinformation manifests in multiple forms, including abnormal user behaviors~\cite{43wang2022gray,44wang2024unveiling,53wang2026graph}, deceptive online reviews~\cite{10shu2017fake,49jiang2025towards,55huang2024eml}, and fabricated news content~\cite{11zhou2020survey,50jiang2025epidemiology,56yin2025emulating}.  
Early research relied heavily on feature engineering and manual rule construction, focusing on linguistic cues, sentiment markers, or propagation patterns to distinguish truthful from deceptive content~\cite{12ott2011finding}.  
With the rise of deep learning, neural architectures such as CNNs, RNNs, and GCNs~\cite{48gao2025graph,52yang2024graph} became dominant, learning high-dimensional semantic and structural representations directly from text or graph-structured data~\cite{13ruchansky2017csi,14wu2023adversarial,51gao2025rethinking}.  
Recent advances further explore hybrid methods, such as PU learning for streaming reviews~\cite{15wu2018hpsd} or graph-based denoising frameworks for fake news~\cite{02yin2025graph}.  
Despite these progressions, misinformation detection has largely been studied through single-model or monolithic paradigms, and to date no work has systematically investigated the use of multi-agent collaboration for this domain.  
This leaves open the question of whether multiple agents, each specialized in different perspectives, can more effectively capture the multifaceted nature of deception.

\noindent\textbf{Multi-Agent Collaboration with LLMs.}
Recent research has highlighted the promise of multi-agent collaboration, showing that a team of LLM-based agents can outperform individual ones in reasoning, planning, and decision-making tasks~\cite{16park2023generative}. Various topologies have been explored to structure inter-agent communication.  
Sequential chain-based pipelines allow information to be passed step by step (e.g., ChatDev~\cite{18qian2024chatdev}, MetaGPT~\cite{19hong2024metagpt}).  
Centralized star structures rely on a hub agent to coordinate subordinates (e.g., AutoGen~\cite{20wu2024autogen}).  
Hierarchical tree organizations introduce multiple layers of coordination (e.g., SoA~\cite{21ishibashi2024self}).  
Graph-based structures employ fully connected or random edges to maximize information sharing, while more recent frameworks such as GPTSwarm~\cite{23zhuge2024gptswarm} and DyLAN~\cite{07liu2024dynamic} explicitly encode agent interactions in layered organizations where different levels play specialized roles.

These designs demonstrate the potential of collective intelligence but remain limited when applied to misinformation detection.  
The key challenge is that misinformation is deliberately deceptive: no single perspective can reliably capture all its forms.  
Effective detection thus requires diverse perspectives and systematic aggregation mechanisms.  
Existing topologies rarely enforce perspective diversity at the agent level or provide structured aggregation across agents.  
Moreover, they lack adaptive structural mechanisms to balance robustness with efficiency, a critical requirement when auditing massive streams of content.  
These gaps motivate the need for a new framework that enforces role differentiation, adapts its structure dynamically, and consolidates heterogeneous perspectives into reliable judgments.

\section{Conclusion}
In this paper, we presented PAMAS, a hierarchical self-adaptive perspective aggregation system that brings LLM-based multi-agent collaboration into misinformation detection. By explicitly differentiating the roles of Auditors, Coordinators, and a Decision-Maker, PAMAS addresses the dual challenge of input-level and structural information drowning, ensuring that weak anomalies are amplified and preserved through progressive judgment aggregation. Beyond robustness, the framework enhances efficiency through adaptive mechanisms that simplify the structure and strengthen agent capabilities during both optimization and inference. Our extensive experiments confirm that PAMAS delivers consistent improvements over strong baselines, advancing not only accuracy but also efficiency and interpretability.

Looking forward, several directions remain open. First, extending PAMAS to handle multimodal misinformation would broaden its applicability to real-world platforms. Second, integrating adaptive human-in-the-loop strategies could further enhance accountability and trust in high-stakes auditing scenarios.

\begin{acks}
This work is supported by the National Natural Science Foundation of China (Grant No. 62176028), the Natural Science Foundation of Chongqing, China (Grant No.CSTB2024NSCQ-MSX0617), the Program of China Scholarship Council (Grant No. 202406050164), the Australian Research Council (Grant No. DE230101033, DE250100613, LP230200892, and DP240101814),  Project to Attract Foreign Experts (Grant No. H20251055), and the Royal Society International Exchange Grant (IEC\textbackslash
NSFC\textbackslash
242347) through Royal Society and NSFC. Coauthor Kai Shu consulted on this paper on unpaid weekends for personal interest.  
\end{acks}

\bibliographystyle{ACM-Reference-Format}
\bibliography{sample-base}


\appendix

\section{Experimental Setting}
\label{appendix:setting}
\noindent\textbf{Datasets.} We use three commonly used public benchmark datasets in our experiments: Amazon~\cite{01wang2022ada}, DeRev2018\footnote{\url{https://fornaciari.netlify.app/dataset/derev18/}}, and PolitiFact~\cite{02yin2025graph}. 
The Amazon dataset represents the e-commerce scenario, where the task is to detect fake or abnormal user accounts that engage in manipulative behaviors such as spam posting or coordinated review injection.  
DeRev2018 is a product reviews dataset with ground-truth veracity labels, specifically designed for deceptive review detection.  
PolitiFact is a widely used fact-checking dataset that focuses on fake news and political misinformation labeled by professional journalists.
The dataset statistics are presented in Table~\ref{datasets}.

\begin{table}[h]
\caption{Statistics of datasets.}
\footnotesize
\label{datasets}
\centering
\begin{tabular}{@{}ccccc@{}}
\toprule
\textbf{Datasets}           & \textbf{\#All} & \textbf{\#True} & \textbf{\#False} & \textbf{Domain} \\ \midrule
\textbf{Amazon}         & 4,904           & 2,989           & 1,915                 & Abnormal User           \\
\midrule
\textbf{DeRev2018}      & 1,552           & 776          & 776               & Misleading Review               \\
\midrule
\textbf{PolitiFact}      & 670          & 348           & 322               & Fake News           \\
\bottomrule
\end{tabular}
\end{table}
\noindent\textbf{Baselines.}  
We compare PAMAS against two categories of baselines: multi-agent collaboration topologies, and representative deep learning and rule-based methods. For Multi-agent baselines:  
\begin{itemize}[leftmargin=12pt]
    \item Vanilla-LLM~\cite{47liu2024deepseek}: A direct LLM API invocation to detect misinformation without any agent structure.
    \item Vanilla-Agent~\cite{46mohammadi2025evaluation}: A single agent with the simplest profile and action memory, serving as the minimal baseline.  
    \item Chain, Star, Tree~\cite{05qian2024scaling}: Classical collaboration structures where agents are connected sequentially (chain), through a central hub (star), or hierarchically (tree).  
    \item Graph~\cite{06zhang2025g}: A fully connected graph where each node is an agent and all pairs exchange information.  
    \item Layer~\cite{23zhuge2024gptswarm}: A fully connected layered structure, where each layer contains agents with different roles.  
    \item DyLAN~\cite{07liu2024dynamic}: A temporal layered design where agents evaluate each other and iteratively update mutual trust weights.  
\end{itemize}

We also compare with representative neural baselines:  
\begin{itemize}[leftmargin=12pt] 
    \item MLP~\cite{09taud2017multilayer}: A standard multilayer perceptron model for detection.  
    \item NFGCN-TIA (For Abnormal User)~\cite{03wang2022detecting}: A graph convolutional model that highlights suspicious user interactions through co-rating relations.  
    \item SIPUL (Misleading Review)~\cite{04shunxiang2023building}: A sentiment-intensity-guided positive–unlabeled learning framework for review detection.  
    \item BREAK (Fake News)~\cite{02yin2025graph}: A bi-level semantic modeling framework that denoises and integrates structural and feature semantics for fake news detection.  
\end{itemize}

\noindent\textbf{Basic Settings.} 
We divide the datasets into three parts: training set, validation set, and test set, following a 7:1:2 ratio. For commonly used evaluation metrics, Accuracy, Precision, Recall, F1, and AUC, are employed. Each metric is calculated 5 times, and the average results are reported. Besides, we implement PAMAS using PyTorch. The backbone LLMs are drawn from three representative models: DeepseekV3.1-Chat\footnote{\url{https://platform.deepseek.com/}}, GPT-5 mini\footnote{\url{https://platform.openai.com/docs/models/gpt-5-mini}}, and Qwen3-8B\footnote{\url{https://huggingface.co/Qwen/Qwen3-8B}}.  
For all baseline multi-agent methods, unless otherwise specified, we use DeepseekV3.1-Chat as the default backbone with 32 agents.  
For PAMAS, we construct a hierarchy of 32 agents, consisting of 16 Auditors, 6 first-level, 5 second-level, 4 third-level Coordinators, and 1 Decision-Maker. For Auditors and Coordinators, each agent randomly selects one model from the three backbone options, introducing diversity at the foundation level. This design aligns with our core objective of enhancing perspective diversity for reliable misinformation detection.  
Unless otherwise stated, the Decision-Maker defaults to DeepseekV3.1-Chat.

\section{Detailed Prompts of Three Agents}
\label{prompts}

\begin{tcolorbox}[
enhanced, colback=gray!10, colframe=gray,
width=\linewidth, arc=1mm, auto outer arc,
title={Auditor Prompts}, breakable, boxsep=0mm,
top=2mm, bottom=2mm, fontupper=\linespread{0.9}\selectfont
]
\textbf{Reasoning Prompt $(\phi_{A})$:}
You are a detection expert. Given user's \textcolor{red}{<Sub-Features>} and your own \textcolor{red}{<Experience>}, decide if the user is malicious (1) or normal (0). Output is \textcolor{Blue}{<User, Decision, Reason>}.

\textbf{Refine Prompt $(\operatorname{SelfRef}_{A})$:}
You are a detection expert. Based on your \textcolor{red}{<Current Experience>}. You have now studied \textcolor{red}{<Other} \textcolor{red}{Agent’s Experience>}. Please integrate both by summarizing them into one concise sentence as your updated experience. Output is \textcolor{Blue}{<New Experience>}.
\end{tcolorbox}

\begin{tcolorbox}[
enhanced, colback=gray!10, colframe=gray,
width=\linewidth, arc=1mm, auto outer arc,
title={Coordinator Prompts}, breakable, boxsep=0mm,
top=2mm, bottom=2mm, fontupper=\linespread{0.9}\selectfont
]
\textbf{Refine Prompt $(\operatorname{SelfRef}_{C})$:}
You are a team coordinator. Your current weight assignments for the members are \textcolor{red}{<Current Weights>}. Based on their historical \textcolor{red}{<Decisions>} and \textcolor{red}{<Reasons>}, slightly adjust each subordinate’s weight. Output is \textcolor{Blue}{<New Weights>}.
\end{tcolorbox}

\begin{tcolorbox}[
enhanced, colback=gray!10, colframe=gray,
width=\linewidth, arc=1mm, auto outer arc,
title={Decision-Maker Prompts}, breakable, boxsep=0mm,
top=2mm, bottom=2mm, fontupper=\linespread{0.9}\selectfont
]
\textbf{Reasoning Prompt $(\phi_{D})$:}  
You are the final decision expert. Given the user’s \textcolor{red}{<Full Features>} and all \textcolor{red}{<Coordinators’ Judgments>} together with your own \textcolor{red}{<Experience>}, decide if the user is malicious (1) or normal (0). Output is \textcolor{Blue}{<User, Decision, Reason>}.  

\textbf{Refine Prompt $(\operatorname{SelfRef}_{D})$:}  
You are the final decision expert. Your current experience is \textcolor{red}{<Current Experience>}. You have now received \textcolor{red}{<Coordinators’ Judgments and Reasons>}. Please integrate both, adjust each coordinator’s weight slightly, and summarize your updated experience in one concise sentence. Output is \textcolor{Blue}{<New Weights, New Experience>}.
\end{tcolorbox}

\section{Algorithm}
The algorithm of PAMAS is referred to as Algorithm \ref{alg:hpamas}.
\begin{algorithm}[t]
\caption{Optimization and  Inference of PAMAS}
\label{alg:hpamas}
\KwIn{Training data $\{(x,y)\}$, epochs $Epochs$, maximum group size $n_{\max}$}
\KwOut{Optimized MAS structure (agent hierarchy), agent memories, confidence weights, and detection outputs (predicted labels)}

\BlankLine
\textbf{Initialization:}\;
Initialize memories as empty and confidence weights to $1$\;
Decision-Maker builds initial $\mathcal{M}^{\text{exp}}_D$ via self-learning\;
Assign Auditors with profile-specific subsets\; 
init $\mathcal{M}^{\text{exp}}_A$ via RAS\;
Build hierarchy via clustering with anchor selection and group expansion (limited by $n_{\max}$)\;

\BlankLine
\For{$\text{epoch}=1$ \KwTo $Epochs$}{
  \textbf{Forward pass:}\;
  Auditors produce $(d_A,r_A)$\; Coordinators aggregate\; Decision-Maker returns $(d_D,r_D)$\;

  \BlankLine
  \textbf{Topology adaptation:}\;
  Prune redundant subordinates\; Expand complementary subordinates\;

  \BlankLine
  \textbf{Backward targeted correction (errors only):}\;
  Decision-Maker: $\mathcal{M}^{\text{exp}}_D \leftarrow \mathcal{M}^{\text{exp}}_D \cup \operatorname{SelfRef}_D(\cdot)$; update $\mathcal{M}^{\text{conf}}_D$\;
  Coordinators: update $\mathcal{M}^{\text{conf}}$ for their subordinates\;
  Auditors: retrieve from $\mathcal{M}^{\text{exp}}_D$ using existing fragments and update $\mathcal{M}^{\text{exp}}_A$\;
}

\BlankLine
\textbf{Inference (confidence-guided routing):}\;
Each Coordinator activates subordinates by descending confidence until a majority emerges, yielding $d_C$\;
Decision-Maker routes over top-level Coordinators, forms $\mathcal{A}^\star$, and integrates $\{(d_{C_k},\alpha_k)\}_{k\in\mathcal{A}^\star_D}$ with $\mathcal{M}^{\text{exp}}_D$ and $\mathcal{F}(x)$ to obtain $(d_D,r_D)$\;
\end{algorithm}

\section{Dimensions of Characteristics}
\label{Characteristic}

To enable perspective diversity in PAMAS, we design 20 characteristic dimensions for each dataset. These dimensions cover two main categories: (1) \textbf{statistical features}, which capture distributional or behavioral patterns such as ratings and activity frequencies, and (2) \textbf{textual semantics}, which capture linguistic properties such as sentiment, style, and consistency. Each Auditor’s profile $\mathcal{P}_A$ specifies only a subset of these dimensions, ensuring complementary perspectives.

For abnormal user detection:
\begin{itemize}[leftmargin=*]
    \item \textbf{Statistical features:} average rating, standard deviation of ratings, rating variance, number of interactions/reviews, model prediction score, maximum rating, minimum rating, median rating, rating range, ratio of positive ratings, ratio of negative ratings, most common rating, average review length, standard deviation of review lengths, maximum review length, minimum review length, median review length, review-length range.
    
    \item \textbf{Textual semantics:} semantic consistency, sentiment summary, opinion diversity, informativeness, detail level, persuasive strength, emotional intensity, subjectivity, bias, readability.
\end{itemize}

For misleading review detection:
\begin{itemize}[leftmargin=*]
    \item \textbf{Statistical features:} deviation from product average rating, ratio of positive to negative helpful votes,  model prediction score, mean helpfulness per review, variance of review helpfulness across a product, long-tail review frequency, duplicate headline ratio, review length variance, median helpfulness per user, outlier score in rating distribution, helpfulness-adjusted rating deviation.
    \item \textbf{Textual semantics:} sentiment polarity, exaggerated phrase frequency, promotional keyword ratio,  contradiction between headline and body, stance divergence from majority reviews,  readability index, review coherence score, ratio of vague adjectives (“good”, “nice”), overuse of comparative/superlative forms.
\end{itemize}

For fake news detection:
\begin{itemize}[leftmargin=*]
    \item \textbf{Statistical features:} article and headline length, model prediction score, headline-to-body length ratio, ratio of unique entities, entity repetition count, external link count, average sentence length, variance of paragraph lengths, ratio of quotations, abnormal frequency of sensational words, cross-source redundancy indicator.
    \item \textbf{Textual semantics:} sentiment polarity, sentiment intensity, subjectivity score, stance inconsistency between headline and body, argument coherence score, lexical diversity, frequency of hedging terms (“may”, “might”), frequency of extreme terms (“always”, “never”), use of clickbait patterns, readability score, claim verifiability score.
\end{itemize}

\section{Analysis of PAMAS}
\label{analysis}
This section analyzes the advantages of PAMAS over existing approaches from two perspectives: interpretability and efficiency.

\begin{table}[h]
\centering
\caption{Comparison of token efficiency across different multi-agent topologies. 
$T$ = token cost per agent call, $N$ = batch size, $p_{\text{err}}$ = error rate, 
$k$ = number of activated agents with routing. 
$n_A$ and $n_C$ denote the number of Auditors and Coordinators in PAMAS. $n$ denotes the number of agents in the general topologies.
For layered structures, $n_\ell$ denotes the number of agents in layer $\ell$.
TA, TC, and CR stand for topology adaptation, targeted correction, and confidence-guided routing.}
\label{tab:efficiency}
\resizebox{0.45\textwidth}{!}{
\begin{tabular}{p{2.4cm}p{2.4cm}p{2.4cm}}
\toprule
\textbf{Topology} & \textbf{Training Cost} & \textbf{Inference Cost} \\
\midrule
Chain & $2N \cdot n \cdot T$ & $n \cdot T$ \\
Star & $2N \cdot n \cdot T$ & $n \cdot T$ \\
Fully connected & $2N \cdot n^2 \cdot T$ & $n^2 \cdot T$ \\
Layered & $2N \cdot (\sum_\ell n_\ell) \cdot T$ & $(\sum_\ell n_\ell) \cdot T$ \\
Tree & $2N \cdot n \cdot T$ & $n \cdot T$ \\
\makecell[l]{PAMAS with\\TA} & 
\makecell[l]{$N \cdot \big[(\tilde{n}_A+1)$ \\ $+ (\tilde{n}_A+\tilde{n}_C)\big]\cdot T$} & $(\tilde{n}_A+1)\cdot T$ \\
\makecell[l]{PAMAS with \\TA+TC} & 
\makecell[l]{$N \cdot \big[(\tilde{n}_A+1)$ \\ $+ p_{\text{err}}(\tilde{n}_A+\tilde{n}_C)\big]\cdot T$} & $(\tilde{n}_A+1)\cdot T$ \\
\makecell[l]{PAMAS with\\TA+TC+CR} & 
\makecell[l]{$N \cdot \big[(\tilde{n}_A+1)$ \\ $+ p_{\text{err}}(\tilde{n}_A+\tilde{n}_C)\big]\cdot T$} & $m \cdot T,\ \ m \ll (\tilde{n}_A+1)$ \\
\bottomrule
\end{tabular}}
\end{table}
 
\subsection{Traceable Interpretability}
Beyond robustness, PAMAS provides \textit{human-friendly interpretability}.  
The hierarchical structure ensures that every decision is traceable back to its rationale.  
At the top level, the Decision-Maker outputs an explanation $r_D$, which summarizes the rationale underlying the final decision.  
If deeper inspection is required, human auditors can trace this explanation downward: Coordinators reveal intermediate rationales aggregated from their subordinates, while Auditors expose information-specific insights such as sentiment shifts or statistical irregularities.  
This strict traceability enables immediate identification of which agent contributed to a judgment and why, ensuring accountability and supporting human–AI collaboration.  

In contrast, existing MAS (chain, star, or fully connected) suffer from either overly entangled communication, making responsibility attribution difficult, or lack of a strict hierarchy, which obscures accountability.  
Similarly, deep learning models typically require post-hoc explanation methods that are less faithful to the actual decision process.  

\subsection{Efficiency in Training and Inference}
PAMAS achieves efficiency gains in both training and inference, mainly by reducing token consumption in LLM-based agents.  

\subsubsection{Training Efficiency.}  
Let $N$ denote the batch size, $p_{\text{err}}$ the error rate, and $n_A$, $n_C$ the numbers of Auditors and Coordinators, respectively.  
If each agent call requires $T$ tokens, then without topology adaptation and targeted correction, the training cost consists of two parts:  
forward inference, where all $n_A$ auditors and the Decision-Maker participate, costing $N \cdot (n_A+1)\cdot T$,  
and refinement, where all $n_A+n_C$ agents update their memories, costing $N \cdot (n_A+n_C)\cdot T$.  
Thus the total cost is $N \cdot \big[(n_A+1) + (n_A+n_C)\big]\cdot T$.  

With topology adaptation, the system prunes redundant agents, reducing the number of auditors and coordinators from $(n_A, n_C)$ to $(\tilde{n}_A, \tilde{n}_C)$, and introduces new ones only when correlated failures emerge. Moreover, with targeted correction, only misclassified samples trigger updates.
The forward cost remains $N \cdot (\tilde{n}_A + 1) \cdot T$,
while the refinement cost reduces to $N \cdot p_{\text{err}} \cdot (\tilde{n}_A + \tilde{n}_C) \cdot T$,
where typically $p_{\text{err}} \ll 1$, further balancing efficiency and robustness. 

\subsubsection{Inference Efficiency.}  
If every inference activates all agents, the token cost is dominated by Auditors and the Decision-Maker, yielding $(\tilde{n}_A+1)\cdot T$.  
Coordinators do not require LLM calls during inference since they only read from the confidence memory.  
PAMAS instead employs \textit{confidence-guided routing}, which activates only the most confident branches until a majority consensus is reached.  
If the decision is resolved after $k \ll (\tilde{n}_A+1)$ activations, the cost reduces to $k \cdot T$,  
resulting in substantial token savings without compromising accuracy.  

\subsubsection{Comparison with Other MAS Topologies.}  
Other multi-agent structures, such as chain, star, fully connected, layered, and tree architectures, all require different token costs depending on their communication pattern.  
Table~\ref{tab:efficiency} summarizes the training and inference costs across representative topologies.  
It can be seen that PAMAS, when equipped with topology adaptation, targeted correction and confidence-guided routing, achieves substantially lower costs than all baselines.

\end{document}